\documentclass[aps,pre,showpacs,twocolumn,showpacs,superscriptaddress,amsmath,groupedaddress]{revtex4}

\usepackage{graphicx,amssymb,bbm,bm}

\begin{document}

\newcommand{\be}   {\begin{equation}}
\newcommand{\ee}   {\end{equation}}
\newcommand{\ba}   {\begin{eqnarray}}
\newcommand{\ea}   {\end{eqnarray}}
\newcommand{\tr}   {{\rm tr}}
\newcommand{\ZZ}{\mathbb{Z}}
\newcommand{\RR}{\mathbb{R}}
\newcommand{\CC}{\mathbb{C}}
\newcommand{\dd} {{\rm d}}

\title{Wigner separability entropy and complexity of quantum dynamics} 

\author{Giuliano Benenti}
\email{giuliano.benenti@uninsubria.it}
\affiliation{CNISM, CNR-INFM \& Center for Nonlinear and Complex Systems,
Universit\`a degli Studi dell'Insubria, Via Valleggio 11, 22100 Como, Italy}
\affiliation{Istituto Nazionale di Fisica Nucleare, Sezione di Milano,
via Celoria 16, 20133 Milano, Italy}
\author{Gabriel G. Carlo}
\email{carlo@tandar.cnea.gov.ar}
\affiliation{ Departamento de F\'{\i}sica, 
              Comisi\'on Nacional de Energ\'{\i}a 
              At\'omica, 
              Avenida del Libertador 8250, 
             (1429) Buenos Aires, Argentina}
\author{Toma\v z Prosen}
\email{tomaz.prosen@fmf.uni-lj.si}
\affiliation{Department of Physics, Faculty of Mathematics and Physics, 
University of Ljubljana, Ljubljana, Slovenia.}

\date{\today}

\begin{abstract}
We propose the Wigner separability entropy as a measure of complexity 
of a quantum state. This quantity measures the number of terms 
that effectively contribute to the Schmidt decomposition of 
the Wigner function with respect to a chosen phase space decomposition.
We prove that the Wigner separability entropy is equal to the 
operator space entanglement entropy, measuring entanglement in the
space of operators, and, for pure states, to twice the entropy
of entanglement. The quantum to classical correspondence 
between the Wigner separability entropy and the separability 
entropy of the classical phase space Liouville density is
illustrated by means of numerical simulations of chaotic maps.
In this way, the separability entropy emerges as an extremely 
broad complexity quantifier in both the classical and quantum realms.
\end{abstract}



\maketitle

\section{Introduction}
\label{sec1}

Measuring complexity in a simple and unified way has been a major and long 
quest in both quantum and classical dynamics. 
While there exists a direct connection between chaos and 
algorithmic complexity of trajectories 
in classical physics~\cite{ford}, 
the problem is particularly elusive for quantum 
mechanics~\cite{alicki}, where 
the notion of trajectory is forbidden by 
the Heisenberg uncertainty principle and 
complexity can be attributed not only to the lack of integrability 
but also to the tensor-product structure of the Hilbert space,
that is, to entanglement.

The phase space representation of quantum mechanics is a 
very convenient framework to investigate quantum complexity, 
in that one can compare classical and quantum dynamical
evolutions of distributions in phase space.
In this context, the number of Fourier harmonics of the Wigner function
has already been used to characterize the complexity of a quantum
state, both for single-particle~\cite{harmonics} 
and many-body~\cite{vinitha} quantum dynamics,
in particular to detect quantum phase transitions. 
This complexity measure can be equally applied to 
classical and quantum mechanics, with the Liouville density 
used instead of the Wigner function in the classical case.  
However, this quantity has the disadvantage of being 
basis-dependent. Moreover, knowledge of the whole 
Fourier harmonics spectrum of the Wigner function seems
in general not necessary to compute expectation values 
of physically relevant observables. 
For instance, even though the number of 
harmonics grows exponentially in time for both integrable
and non-integrable quantum chaotic Ising chains~\cite{vinitha}, 
the resources required to simulate both local and extensive 
observables grow exponentially in the chaotic case but 
only polynomially for the integrable model~\cite{prosenznidaric}. 

Recently, a new complexity indicator has 
been introduced for classical dynamics, {\em the separability 
entropy}~\cite{prosen1}, measuring the logarithm of the effective
number of terms in the Schmidt (or singular value) decomposition of the 
Liouville density, with respect to an arbitrary phase space 
decomposition. This quantity estimates the minimal amount of 
computational resources required to simulate the classical Liouvillian
evolution and grows linearly in time for dynamics
that cannot be efficiently simulated.
In this paper, we extend this notion of complexity to the quantum 
realm, by defining the Wigner separability entropy as the number
of terms that effectively contribute to the Schmidt decomposition
of the Wigner function.
We prove that such quantity is equal to the operator space 
entanglement entropy~\cite{prosenOSEE}, constructed from the 
Schmidt decomposition of the density operator 
in the space of Hilbert-Schmidt operators and quantifying
the complexity of time-dependent density-matrix renormalization
group simulations.
Furthermore, for pure states 
the Wigner separability entropy is twice the entanglement
entropy. 
We illustrate the quantum to classical correspondence for the 
separability entropy and its link to the entanglement entropy
by means of numerical simulations of quantum chaotic maps. 

The paper is organized as follows. 
In Sec.~\ref{sec2} we define the 
Wigner separability entropy $h[W]$ and prove its relation 
to the operator space entanglement entropy and, for 
pure states, to the entanglement entropy. 
The dynamical evolution of $h[W]$ is studied for quantum chaotic
map models in Sec.~\ref{sec3}. We finish with concluding 
remarks in Sec.~\ref{sec4}.

\section{Wigner separability entropy}
\label{sec2}

Given a system described in a $2d$-dimensional compact phase space 
$\Omega$ by the Wigner 
function $W({\boldsymbol z})$ (with the normalization constraint
$\int\dd{\boldsymbol z} W({\boldsymbol z})=1$) 
and an arbitrary phase space decomposition,
$\Omega=\Omega_1 \oplus \Omega_2$, into 
two set of coordinates, ${\boldsymbol z}\equiv ({\boldsymbol x},{
\boldsymbol y})$, we can write
the Schmidt (singular value) decomposition of the Wigner function:
\begin{equation}
W({\boldsymbol x},{\boldsymbol y})=\sum_n \mu_n 
v_n({\boldsymbol x}) w_n({\boldsymbol y}),
\label{eq:SVDWigner}
\end{equation}
with $n\in\mathbb{N}$, $\{v_n\}$ and $\{w_n\}$ orthonormal 
bases for $L^2(\Omega_1)$ and $L^2(\Omega_2)$, respectively, 
and the Schmidt coefficients 
(singular values) $\mu_1\ge \mu_2\ge \ldots \ge 0$
satisfying $\sum_n \mu_n^2=\int\dd{\boldsymbol z} 
W^2({\boldsymbol z})$.
We then define the \emph{Wigner separability entropy} as 
\begin{equation}
h[W]=-\sum_n \tilde{\mu}_n^2 \ln \tilde{\mu}_n^2,
\label{eq:Wignerentropy}
\end{equation}
where 
\begin{equation}
\tilde{\mu}_n\equiv 
\frac{\mu_n}{\sqrt{\int\dd{\boldsymbol z} W^2({\boldsymbol z})}}.
\end{equation}
The coefficients $\{\tilde{\mu}_n\}$ satisfy 
$\sum_n \tilde{\mu}_n^2=1$ and are the Schmidt coefficients 
of the singular value decomposition of 
$\tilde{W}\equiv W/\sqrt{\int\dd{\boldsymbol z} W^2({\boldsymbol z})}$.
That is, $\tilde{W}$ is normalized in $L^2(\Omega)$:
$\int\dd {\boldsymbol z} \tilde{W}^2({\boldsymbol z})=1$.
Note that for pure states 
\begin{equation}
\tilde{W}({\boldsymbol z}) =(2\pi\hbar)^{-d/2} W({\boldsymbol z}).
\end{equation}
The Wigner separability entropy $h[W]$ quantifies the logarithm of the
number of terms that effectively contribute to decomposition 
(\ref{eq:SVDWigner}) and therefore provides a 
measure of separability of the Wigner function with respect 
to the chosen phase space decomposition. 

The main advantage of defining the separability entropy 
in phase space by means of the Wigner function is that such 
quantity can be directly translated to classical mechanics. 
The classical analog of the Wigner separability entropy is 
the s-entropy 
$h[\rho_c]$ defined in Ref.~\cite{prosen1}, where the classical
phase space distribution $\rho_c({\boldsymbol z})$ is used instead of the
Wigner function $W({\boldsymbol z})$. 

It is interesting to establish a connection between the Wigner 
separability entropy and the 
\emph{operator space entanglement entropy}~\cite{prosenOSEE}, 
constructed from the 
Schmidt decomposition of the density operator $\hat{\rho}$
acting on the Hilbert space $\mathcal{H}$.
Since $\text{Tr}(\hat{\rho}^2)\le 1$, the density operator is a  
Hilbert-Schmidt operator, that is, an operator with finite
Hilbert-Schmidt norm $\|\hat{\rho}\|_{\rm HS}=\sqrt{\text{Tr}(\hat{\rho}^2)}$.
Therefore, $\hat{\rho}$ can be interpreted as a vector belonging
to the Hilbert space $B(\mathcal{H})$ of Hilbert-Schmidt operators,
with the Hilbert-Schmidt inner product 
$\langle \hat{A}, \hat{B} \rangle_{\rm HS}=\text{Tr}
(\hat{A}^\dagger \hat{B})$.
Therefore, given $\mathcal{H}=\mathcal{H}_1\otimes\mathcal{H}_2$,
the density operator has a Schmidt decomposition,
\begin{equation}
\hat{\rho}=\sum_n \mu_n \hat{\sigma}_{n}\otimes\hat{\tau}_{n},
\label{SVDrho}
\end{equation}
where $\{\hat{\sigma}_{n}\}$ and $\{\hat{\tau}_{n}\}$ are orthonormal ($\text{Tr}(\hat{\sigma}^\dagger_m \hat{\sigma}_n) = \delta_{m,n}$, $\text{Tr}( \hat{\tau}^\dagger_m \hat{\tau}_n) = \delta_{m,n}$)
bases for $B(\mathcal{H}_1)$ and $B(\mathcal{H}_2)$, respectively, 
and the Schmidt coefficients $\mu_1\ge \mu_2\ge \ldots \ge 0$
satisfying $\sum_n \mu_n^2=\text{Tr}(\hat{\rho}^2)=\|\hat{\rho}\|_{\rm HS}^2$.
The operator space entanglement entropy~\cite{prosenOSEE} is then given by 
\begin{equation}
h[\hat{\rho}]=-\sum_n \tilde{\mu}_n^2 \ln \tilde{\mu}_n^2,
\quad\tilde{\mu}_n\equiv 
\frac{\mu_n}{{\|\hat{\rho}\|_{\rm HS}}}.
\label{eq:OSEE}
\end{equation}

In what follows, we prove that $h[\hat{\rho}]=h[W]$, that is,
the operator space entanglement entropy equals the Wigner 
separability entropy. This result follows from the fact that the 
Weyl correspondence establishes an isomorphism between 
Hilbert-Schmidt operators and $L^2(\Omega)$ functions on classical phase 
space~\cite{pool,folland}. Since the density operator $\hat{\rho}$ 
is the integral kernel of a unique linear one-to-one transformation mapping the
operators $\hat{\sigma}_n \leftrightarrow \hat{\tau}_n$, $\forall n$,
i.e. $\text{Tr}_1((\hat{\sigma}^{\dagger}_{n})_1 \hat{\rho})=\mu_n \hat{\tau}_{n}$, 
and vice versa for the inverse transformation,
$\text{Tr}_2((\hat{\tau}^{\dagger}_{n})_2 \hat{\rho})=\mu_n \hat{\sigma}_{n}$, 
and since the Weyl transform of the density operator is the 
Wigner function, it follows that $\hat{\rho}$ and $W$ have the 
same Schmidt coefficients, and therefore 
\begin{equation}
h[\hat{\rho}]=h[W].
\label{eq:rhoW}
\end{equation} 

In order to explicitly illustrate the above result, we consider a
density operator 
$\hat{\rho}(\hat{a_1},...,\hat{a}_d,
\hat{a}_1^\dagger,...,\hat{a}_d^\dagger)$ written in terms of a set of 
bosonic creation-annihilation operators, 
$[\hat{a}_i,\hat{a}_j]=
[\hat{a}_i^\dagger,\hat{a}_j^\dagger]=0$,
$[\hat{a}_i^\dagger,\hat{a}_j]=\delta_{ij}$, and define
the Wigner function as
\begin{eqnarray}
W({\boldsymbol\alpha},{\boldsymbol\alpha}^{*})=\frac{1}{\pi^{2d}
\hbar^d} && \int\dd{\boldsymbol\eta}\dd{\boldsymbol\eta}^*
\exp{\left(\frac{{\boldsymbol\eta}^*\cdot{\boldsymbol\alpha}}{\sqrt{\hbar}}
-\frac{{\boldsymbol\eta}
\cdot{\boldsymbol\alpha}^*}{\sqrt{\hbar}}\right)} \nonumber\\
&&\times \text{Tr}[
\hat{\rho}\hat{D}({\boldsymbol\eta})],
\label{eq:wignerD}
\end{eqnarray}
where ${\boldsymbol\eta}=(\eta_1,...,\eta_d)\in\CC^d$ and
${\boldsymbol\alpha}=(\alpha_1,...,\alpha_d)\in\CC^d$ are $d$-dimensional
complex variables, the integration runs over the complex
$\eta_i$-planes for $i=1,...,d$,  $\hat{D}$ is the displacement operator
\begin{equation}
\hat{D}\left({\boldsymbol\eta}\right)
=\hat{D}\left(\eta_1,...,\eta_d\right)=
\exp\left[\sum_{i=1}^{d}{\left(\eta_i\hat{a_{i}}^
{\dagger}-\eta_i^{*}\hat{a_{i}}\right)}\right],
\end{equation}
and $|{\boldsymbol\alpha}\rangle$ are the coherent states
\begin{equation}
|{\boldsymbol\alpha}\rangle=
|\alpha_{1}\alpha_{2}...  \alpha_{d}\rangle=
\hat{D}\left(\frac{{\boldsymbol\alpha}}{\sqrt{\hbar}}\right)|0\rangle,
\end{equation}
with $|\alpha_i\rangle$ being an eigenstate of the annihilation operator
$\hat{a}_i$, i.e., $\hat{a}_i|\alpha_i\rangle= \frac{\alpha_i}{\sqrt{\hbar}}
|\alpha_i\rangle$,
and $|0\rangle$ being the vacuum state $\hat{a}_j|0\rangle = 0$.
Using the singular value decomposition
\begin{eqnarray}
\hat{\rho}&=&\sum_n \mu_n 
\hat{\sigma}_n(\hat{a}_1,...,\hat{a}_{d/2},
\hat{a}_1^\dagger,...,\hat{a}_{d/2}^\dagger) \nonumber \\
&& \otimes 
\hat{\tau}_n(\hat{a}_{d/2+1},...,\hat{a}_{d},
\hat{a}_{d/2+1}^\dagger,...,\hat{a}_{d}^\dagger),
\label{eq:SVDrhobosonic}
\end{eqnarray}
we obtain
\begin{eqnarray}
\text{Tr}[
\hat{\rho}\hat{D}({\boldsymbol\eta})]
&=&\sum_n \mu_n
\tilde{\sigma}_n(\eta_1,...,\eta_{d/2},
\eta_1^*,...,\eta_{d/2}^*) \nonumber \\
&&\times\tilde{\tau}_n(\eta_{d/2+1},...,\eta_{d},
\eta_{d/2+1}^*,...,\eta_{d}^*),
\end{eqnarray}
where
\begin{eqnarray}
\tilde{\sigma}_n
& \equiv & \text{Tr}_{1,...,d/2}\,[\hat{\sigma}_n \hat{D}(\eta_1,...,\eta_{d/2})], \\
\tilde{\tau}_n
& \equiv & \text{Tr}_{d/2+1,...,d}\,[\hat{\tau}_n 
\hat{D}(\eta_{d/2+1},...,\eta_{d})].
\end{eqnarray}
Finally, we derive
\begin{eqnarray}
W({\boldsymbol\alpha},{\boldsymbol\alpha}^{*}) &=&
\sum_n \mu_n 
v_n(\alpha_1,...,\alpha_{d/2},
\alpha_1^*,...,\alpha_{d/2}^*) \nonumber \\
&& \times w_n(\alpha_{d/2+1},...,\alpha_{d},
\alpha_{d/2+1}^*,...,\alpha_{d}^*),
\label{SVDW}
\end{eqnarray}
where 
\begin{eqnarray}
v_n &\equiv& \frac{1}{\pi^{d} \hbar^{d/2}}
\int\dd\eta_1\dd\eta^*_1\cdots \dd\eta_{d/2}\dd\eta_{d/2}^* \nonumber \\
&\times&\exp\left[\sum_{i=1}^{d/2} 
\left(\frac{\eta_i^* \alpha_i}{\sqrt{\hbar}}- 
\frac{\eta_i \alpha_i^*}{\sqrt{\hbar}}\right)\right]
\tilde{\sigma}_n,  \label{eq:vn} \\
w_n &\equiv& \frac{1}{\pi^{d} \hbar^{d/2}}
\int\dd\eta_{d/2+1}\dd\eta_{d/2+1}^*\cdots \dd\eta_{d}\dd\eta^*_{d} \nonumber \\
&\times& \exp\left[\sum_{i=d/2+1}^{d} 
\left(\frac{\eta_i^* \alpha_i}{\sqrt{\hbar}}- 
\frac{\eta_i \alpha_i^*}{\sqrt{\hbar}}\right)\right]
\tilde{\tau}_n. \label{eq:wn}
\end{eqnarray}
Eq. (\ref{SVDW}) is precisely the singular value decomposition of the 
Wigner function (the orthogonality of the Schmidt vectors (\ref{eq:vn},\ref{eq:wn}) can be checked straightforwardly)
and has the same Schmidt coefficients $\{\mu_n\}$ as the
the singular value decomposition 
(\ref{eq:SVDrhobosonic}) of the density operator, so we conclude that $h[W]=h[\hat{\rho}]$ (\ref{eq:rhoW}).

When the density operator $\hat{\rho}$ describes a pure state,
$\hat{\rho}=|\psi\rangle\langle\psi|$, there exists a simple relation
between the Wigner separability entropy and the entanglement 
content of the state $|\psi\rangle\in \mathcal{H}=
\mathcal{H}_1\otimes \mathcal{H}_2$. The Schmidt decomposition 
of $|\psi\rangle$ reads~\cite{nielsen,qcbook} 
\begin{equation}
|\psi\rangle=\sum_ j \lambda_j |\phi_j\rangle
\otimes |\xi_j\rangle,
\end{equation}
with $\{\phi_j\}$ ($\{\xi_j\}$) orthonormal basis for 
$\mathcal{H}_1$ ($\mathcal{H}_2$),
$\lambda_1\ge \lambda_2 \ge\ldots \ge 0$,
and $\sum_j \lambda_j^2=1$. 
On the other hand, we can also write the Schmidt decomposition 
of the operator $\hat{\rho}=|\psi\rangle\langle\psi|$:
\begin{equation}
\hat{\rho}=\sum_{j,k} \lambda_j \lambda_k 
|\phi_j\rangle\langle \phi_k|\otimes
|\xi_j\rangle\langle \xi_k|. 
\label{SVDpsi}
\end{equation}
The comparison between (\ref{SVDpsi}) and 
(\ref{SVDrho}) implies 
$\{\mu_n\}_{n\in\mathbb{N}}=
\{\lambda_j\lambda_k\}_{j,k\in\mathbb{N}}$.
Therefore,
\begin{eqnarray}
h[W]&=&-\sum_n \mu_n^2 \ln \mu_n^2=
-\sum_{j,k} \lambda_j^2 \lambda_k^2 \ln
(\lambda_j^2\lambda_k^2) \nonumber\\
&=&-2\sum_{j} \lambda_j^2 \ln \lambda_j^2=-2 
S(\hat{\rho}_1)=-2 S(\hat{\rho}_2),
\end{eqnarray}
where $\hat{\rho}_1=\text{Tr}_2(\hat{\rho})$ and
$\hat{\rho}_2=\text{Tr}_1(\hat{\rho})$
are the reduced density operators 
for subsystems 1 and 2 and $S$ is the von Neumann
entropy. 
Since for a pure state $|\psi\rangle$ von Neumann entropy of the reduced density matrix
quantifies the entanglement $E$ of $|\psi\rangle$~\cite{nielsen,qcbook},
\begin{equation}
E(|\psi\rangle)=S(\hat{\rho}_1)=S(\hat{\rho}_2),
\end{equation}
we can conclude that the Wigner separability entropy is twice
the \emph{entanglement entropy} $E(|\psi\rangle)$: 
\begin{equation}
h[W]=2\,E(|\psi\rangle).
\end{equation}

For pure states $S(\hat{\rho})=0$, and therefore the 
Wigner separability entropy is equal to the 
\textit{quantum mutual information}
\begin{equation}
I(1:2)=S(\hat{\rho}_1)+S(\hat{\rho}_2)-S(\hat{\rho}).
\end{equation}
Quantum mutual information measures total correlations,
both of classical and quantum nature, 
between subsystems 1 and 2 and for pure states 
classical correlations $C(|\psi\rangle)$ 
are equal to quantum correlations, measured by 
$E(|\psi\rangle)$~\cite{vedral}.

\section{Wigner separability entropy for chaotic maps}
\label{sec3}

In order to illustrate the quantum to classical correspondence
for the separability entropy, we study the evolution in time
of $h[W]$ and of its classical counterpart $h[\rho_c]$ for 
three quantum maps: (i) the quantum baker's map, (ii) the perturbed 
quantum cat map, and (iii) a model consisting of two coupled perturbed 
cat maps. The first model (i) exhibits an {\em atypical} failure of 
the quantum to classical correspondence due to discontinuity in the mapping 
producing drastic quantum diffraction effects. The second
model (ii) is {\em typical} in that the Wigner separability entropy 
$h[W]$ follows the classical separability entropy $h[\rho_c]$ up 
to the Ehrenfest time scale $t_{\rm E}$, until Wigner function evolution (whose time derivative is given by the Moyal bracket) is well approximated
by the evolution of the classical Liouville density (whose time derivative is given by the Poisson bracket, the first order in $\hbar$ expansion of the Moyal bracket), i.e.
\begin{equation}
W(t=0)= \rho_c(t=0) \Rightarrow\;
h[W(t)] = h[\rho_c(t)] \; {\rm for} \; t < t_{\rm E}.
\end{equation}
Note that for chaotic dynamics with Lyapunov exponent $\lambda$, the Ehrenfest time scales as $t_{\rm E} \sim (-\log\hbar)/\lambda$ \cite{Zaslavsky}.
The third model (iii) illustrates for 
pure states the connection between the separability entropy and 
the entanglement entropy.

We study the quantum versions of classical 
chaotic maps on the 2-torus $[0,1) \times [0,1)$. Quantization 
imposes the Planck constant to coincide with
the inverse of the dimension $N$ of the Hilbert space, 
i.e., $\hbar = 1/(2 \pi N)$.
The baker map has been studied in detail in many works (see for example 
\cite{Saraceno1, Saraceno2}). The classical baker map is defined by 
\begin{equation}
  (q_{t+1},p_{t+1})=
  \left\{
    \begin{array}{ll}
      (2q_t,\frac12p_t),           & \text{if} \quad 0 \le q_t \le \frac12,
\\
      (2q_t-1,\frac12p_t+\frac12), & \text{if} \quad \frac12 < q_t < 1 ,
    \end{array}
  \right.
  \label{cbaker}
\end{equation}
where the discrete time $t\in\ZZ$ measures the number of 
map iterations.
The quantum analogue of the classical baker map is a unitary 
operator acting on an $N$-dimensional Hilbert space (assuming $N$ to be {\em even}).
In the position($q$)-representation its matrix reads:
\be
B= G_{N}^{-1} \left( \begin{array}{cc}
                           G_{N/2} &      0            \\
                              0    &   G_{N/2}
                          \end{array}
                  \right) \, ,
\label{bakerqq}
\ee
where $G_N$ is the $N$-dimensional antiperiodic Fourier matrix:
\be
(G_{N})_{lj}=
\frac{1}{\sqrt N} \, e^{-2\pi i (j+1/2)(l+1/2)/N} \, ,
\ee
with $0 \le l,j \le N-1$.

The classical perturbed cat map reads \cite{Ozorio}
\begin{equation}
 \left(
    \begin{array}{l}
  q_{t+1} \\ p_{t+1}
  \end{array} \right) =
  \cal{M}
 \left(
    \begin{array}{l}
      q_t           
\\
      p_t + \epsilon(q_t)
    \end{array}
  \right),
  \label{cpcat}
\end{equation}
where $q$ and $p$ are
taken $\mod{(1)}$, and where $\epsilon(q_t)=-[K/(2 \pi)] \sin{(2 \pi q_t)}$. Throughout this paper we have used  
\begin{eqnarray}
\label{eq:ccat}
\cal{M}&=&\left(\begin{array}{cc}2&1\\3&2\end{array}\right),
\end{eqnarray}
satisfying the symplectic condition $\det{\cal M}=1$,
and $K=0.5$.
The perturbed quantum cat map in the $q$-representation 
is given by the $N\times N$ unitary matrix $M$ whose elements read
\begin{equation}
\label{eq:qcat}
{M}_{lj}=  \\ A \,
\exp{\left[\frac{\pi{\rm i}}{N {\cal M}_{12}}( {\cal M}_{11} l^2 - 2lj+ {\cal M}_{22}j^2)
+ F \right]}, 
\end{equation}
where 
\begin{eqnarray*}
A&=& [1/({\rm i} N {\cal M}_{12})]^{1/2}, \\
F&=& [i K N/(2 \pi)] \cos{(2 \pi l /N)}.
\end{eqnarray*}

Finally, we have considered a 2D model (in 4D phase space $(q^1,q^2,p^1,p^2) \in [0,1)^4$) consisting of two coupled perturbed 
cat maps. In order to do this in a symplectic way we have used a symmetric coupling in the position 
coordinates \cite{Wood}. In this case the classical map is given by (superscripts 
indicate which `cat' the coordinates refer to)
\begin{equation} \nonumber
 \left(
    \begin{array}{l}
  q^{1}_{t+1} \\ p^{1}_{t+1} 
  \end{array} \right) =
  \cal{M}
 \left(
    \begin{array}{l}
      q^{1}_t           
\\
      p^{1}_t + \epsilon(q^{1}_t) + \epsilon'(q^{1}_t,q^{2}_t)
    \end{array}
  \right)
\end{equation} and,
\begin{equation}
 \left(
    \begin{array}{l}
  q^{2}_{t+1} \\ p^{2}_{t+1}
  \end{array} \right) =
  \cal{M}
 \left(
    \begin{array}{l}
      q^{2}_t           
\\
      p^{2}_t + \epsilon(q^{2}_t) + \epsilon'(q^{1}_t,q^{2}_t)
    \end{array}
  \right),
  \label{ccpcat}
\end{equation}
where $q^i$ and $p^i$ ($i=1,2$) are again
taken $\mod{(1)}$, and where we have used the same perturbation $\epsilon(q)$ as before (for both maps).
Here, $\epsilon'(q^{1}_t,q^{2}_t)=-[K_c/(2 \pi)] \sin{(2 \pi q^{1}_t+2 \pi q^{2}_t)}$ is the coupling term.

The quantized version of the 2D perturbed cat map is obtained by multiplying a separable (tensor) product of quantized 1D perturbed cat maps with a simple exponentiated coupling matrix (which is diagonal in the $q$-representation). More explicitly, $N^2 \times N^2$ unitary matrix $M^{2D}$ is given by
\begin{equation}
M^{2D}_{(l^1,l^2),(j^1,j^2)} = M_{l^1,j^1} M_{l^2,j^2} C_{j^1,j^2}
\label{eq:qccat}
\end{equation}
where 
\begin{equation}
C_{j^1,j^2}=\exp\{[i K_c N/(2 \pi)] \cos[(2 \pi /N) (j^1+j^2)]\}
\end{equation} 
and $j^1,j^2,l^1,l^2 \in \{0,\ldots,N-1\}$.
In the following calculations 
we take $K_c=1.0$. 

We have studied the classical and 
quantum separability entropies. In all cases we have used the initial states given in terms of
a Gaussian phase space distribution with dispersion equal to $\sqrt{\hbar}$, and its 
quantum analogue, a coherent state on the torus. Both distributions 
are centered in the middle of the phase space, i.e., at $(q,p)=(0.5,0.5)$. 
We have evolved 10 time steps (iterations) for all maps. 

\begin{figure}[htp]
\vspace{1.0cm}
\includegraphics[angle=0.0, width=8cm]{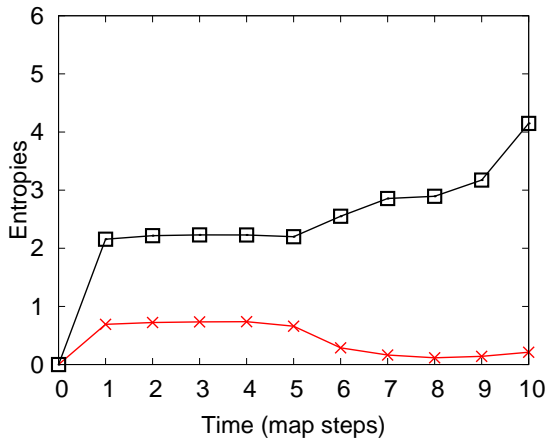}
\caption{(Color online) Classical and quantum separability entropies behavior 
for the baker map. The (red) gray line with crosses corresponds to 
the classical result and the black line with squares to the quantum one.
The quantum curve has been calculated using the Wigner function without 
ghost images~\cite{dittrich}. In both cases $N=2^9$. }
\label{fig1}
\end{figure}

In Fig.~\ref{fig1} we show the separability entropy \cite{prosen1} as a function of 
time (in units of map steps) for the baker map (\ref{cbaker},\ref{bakerqq}). We have used the same discretization number $N=2^9$ for both, the classical and the 
quantum simulations and removed the effects of the torus periodicity on the Wigner distributions 
\cite{dittrich}. We have decomposed the 2-torus in coordinates $q$ and $p$. 
While the classical separability entropy saturates to a value of order $\sim 1$ bit, after an initial short
transient growth, due to the exact solvability of its Liouville
evolution~\cite{gaspard}, and drops back towards zero for $t \sim \log N$ due to coarse-graining of classical Liouvillean evolution, 
the Wigner separability entropy grows with time due to quantum interference and diffraction patterns produced by discontinuity of the map. 
The baker's map is therefore non generic, in that it exhibits 
very different classical and quantum results even within the Ehrenfest time range due to diffraction effects.

%
%

\begin{figure}[htp]
\hspace{0.0cm}
\includegraphics[angle=0.0, width=8cm]{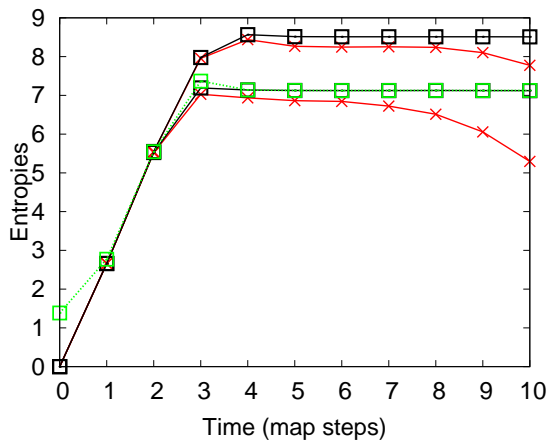}
\caption{(Color online) Same as in Fig.~\ref{fig1} but now for the 
perturbed cat map. The black lines with squares correspond to 
quantum results and the gray (red) ones with crosses to classical ones.  
The light gray (green) dotted line corresponds to the quantum values using the Wigner distribution 
with ghost images which exhibits small deviations at shorter times. 
The lower (three) curves correspond to $N=2^{11}$ while the upper (two) curves correspond to $N=2^{13}.$
}
\label{fig2}
\end{figure}

In order to investigate a case with generic behavior
 we have taken the smooth (continuous) perturbed cat map (\ref{eq:ccat},\ref{eq:qcat}). In Fig.~\ref{fig2} we show 
the same quantities as in Fig.~\ref{fig1}. Now there is quantum to classical 
correspondence, up to the Ehrenfest time $t_{\rm E} \propto \log N$. Note that, similarly to the case of baker map, the classical result also
exhibits discretization time scale after which the classical distribution experiences a 
complexity reduction which is a numerical artifact due to finiteness of $N$.
Indeed, due to coarse graining the Liouville distribution becomes 
homogeneous in phase space, and this implies asymptotically vanishing separability entropy.
Finally, and for comparison purposes only, 
we show the quantum results without removing the effects of the 
ghost images 
that appear on account of the torus periodicity 
(this is shown just for the lower value of $N$).
Wigner separability entropy without ghost images removed exhibits some deviations for shorter times 
(then, the fact that we have ghost peaks in Wigner distribution is more important than at later times), 
the removal of ghost images is therefore needed in order to obtain 
quantum to classical correspondence.

%
%

\begin{figure}[htp]
\hspace{0.0cm}
\includegraphics[angle=0.0, width=8cm]{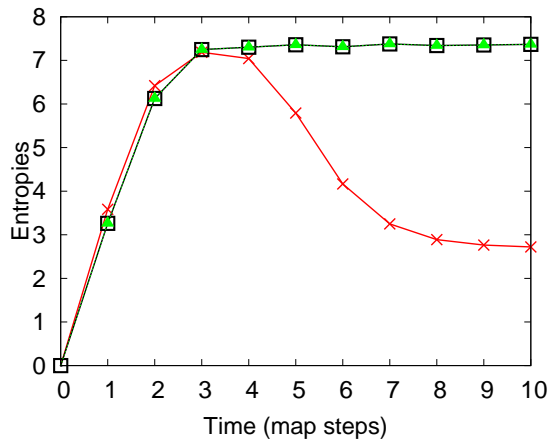}
\caption{(Color online) Classical and quantum separability entropies behavior 
for two coupled cat maps (the gray (red) line with crosses and the black line 
with squares, respectively). Two times von Neumann entropy is also shown by means 
of a light gray (green) dotted line with triangles. We have used discretization number $N=2^6$ in both the 
classical and quantum cases, for each map.}
\label{fig3}
\end{figure}

Finally, we have tested the Wigner separability entropy and its relation to
pure-state entanglement for two coupled  
perturbed cat maps (\ref{ccpcat},\ref{eq:qccat})
(in this case we have evaluated the Wigner distributions according to Ref.~\cite{Konio}). 
In practice we have decomposed the four-dimensional phase space in the 
coordinates $(q^1,p^1)$ 
and $(q^2,p^2)$ corresponding to each of the two `cats', respectively. We have 
also compared the results with the classical separability entropy \cite{prosen1}. 
It is worth mentioning that in order to obtain the 
classical complexity measure we have used the same phase space decomposition  
for the four dimensional Liouville distribution that we have previously  
applied to the four-dimensional Wigner function. 
All these results are shown in Fig.~\ref{fig3}. The agreement between the 
quantum measures is accurate within the machine precision, i.e. $h[W]=2E(|\psi\rangle)$ 
(in the figure we have rescaled von Neumann entropy by a factor of 2), 
reflecting the match between singular 
value decompositions for the density matrix and its Wigner distribution.
In this case we did not remove ghost images in the Wigner distributions  
since all the coherences measured by the separability entropy in these 
coordinate pairs are relevant to measure entanglement. 
Regarding the classical separability entropy we find a reasonable 
agreement up to time $\sim \log N$ where due to coarse graining 
$h[\rho_c]$ starts to drop. It is important to underline that the classical 
saturation is entirely due to the phase space coarse graining chosen for the numerical 
simulations, while the quantum one is an unavoidable phenomenon fixed by 
the finite size of $\hbar$. Note that, as recently pointed 
out~\cite{casati}, 
the initial growth of entanglement can be reproduced in the 
semiclassical regime by purely classical computations. 

%
%
%

\section{Conclusions} 
\label{sec4}

In this paper we have proposed a new measure of complexity
of quantum states, the Wigner separability entropy.
This quantity turns out to be equal to the operator space
entanglement entropy, it quantifies the minimal amount of computational
resources required to simulate the quantum dynamical evolution  
of a system by means of time-dependent density-matrix renormalization 
group \cite{prosenznidaric}. Moreover, due to its relation with the entropy of 
entanglement for pure states and to the existence of the 
analogous s-entropy of Liouville densities in classical dynamics,
the separability entropy emerges as an extremely 
broad complexity quantifier in both the classical and quantum realms.

With regard to a previously proposed phase-space quantum complexity 
indicator, that is, the number of Fourier harmonics of the 
Wigner function~\cite{harmonics,vinitha}, the Wigner separability 
entropy has the advantage of being basis-independent. 
More importantly, numerical
indications~\cite{prosenznidaric} suggest that the Wigner separability 
entropy should be able to distinguish between quantum chaotic and
quantum regular motion also for many-body systems.
Finally, we point out that the Wigner separability entropy 
is well defined also for mixed states and could therefore
be used to quantify the complexity of decoherent quantum 
dynamics.  

\section*{Acknowledgments}

We thank Italo Guarneri for useful discussions.
Partial financial support from Conicet, Argentina (GC), and the grants P1-0044 and J1-2208 of Slovenian Research Agency (TP) is greatfully acknowledged.

\vspace{3pc}


\end{document}